\documentclass[]{webofc}
\usepackage[varg]{txfonts}   
\usepackage{hyperref}
\def\aap{{A\&A}}		
\def\aj{{AJ}}			
\def\apj{{ApJ}}			
\def\apjl{{ApJ}}		
\def\apjs{{ApJS}}		

\def\mnras{{MNRAS}}
\def\nat{{Nature}}
\def\micron{\ensuremath{\mu\textrm{m}}}

\woctitle{Hot Planets and Cool Stars}
\begin{document}
\title{Re-evaluating Hot Jupiter WASP-12b: An Update}
%
%

\author{
Ian J. M. Crossfield\inst{1}\fnsep\thanks{\email{ianc@mpia.de}} \and
Travis Barman\inst{2}\fnsep \and
Brad M. S. Hansen\inst{3}\fnsep \and
Ichi Tanaka\inst{4}\fnsep \and
Tadayuki Kodama\inst{4}\fnsep
}

\institute{Max-Planck Institut f\"ur Astronomie, K\"onigstuhl 17, D-69117, Heidelberg, Germany
\and
           Lowell Observatory, 1400 West Mars Hill Road, Flagstaff, AZ 86001, USA
\and
           Department of Physics \& Astronomy, University of California Los Angeles, Los Angeles, CA 90095, USA
\and
           Subaru Telescope, National Astronomical Observatory of Japan, 650 North A'ohoku Place, Hilo, HI 96720, USA
}

\abstract{%
  The hot Jupiter WASP-12b is one of the largest, hottest, and
  best-studied extrasolar planets.  We revisit our recent analysis of
  WASP-12b's emission spectrum in light of near-infrared spectroscopic
  measurements which have been claimed to support either a
  hydride-dominated or carbon-rich atmospheric composition. We show
  that this new spectrum is still consistent with a featureless
  blackbody, indicating a nearly isothermal photosphere on the
  planet's day side. Thus the ensemble of occultation measurements for
  WASP-12b is still insufficient to constrain the planet's
  atmospheric composition. }
\maketitle

\section{Introduction}
\label{intro}
Although hot Jupiters occur less frequently around sun-like stars than
do less massive planets \citep{howard:2010,howard:2012}, these larger
objects remain the most commonly known type of transiting
planet\footnote{As of 31 Dec 2012: \url{http://www.exoplanet.eu} and
  \url{http:/www.exoplanets.org}} because of the limiting sensitivity
of current ground-based surveys.  In addition, the high temperatures
and large  radii of hot Jupiters make these planets
especially favorable targets for atmospheric characterization via
transits \citep{henry:2000,charbonneau:2000}, secondary eclipses
\citep{charbonneau:2005,deming:2005}, and phase curves
\citep{harrington:2006,knutson:2007b}.  Thus the population of
exoplanets for which atmospheric measurements have been made or
attempted is largely limited to hot Jupiters.

The hot Jupiter WASP-12b is one of the largest and hottest transiting
planets known \citep{hebb:2009,chan:2011,maciejewski:2011}.  The
planet is intensely irradiated by its host star, which gives it an
especially favorable planet/star flux contrast ratio. These conditions have
motivated a flurry of photometric and spectroscopic occultation
measurements
\citep{lopez-morales:2010,croll:2011a,campo:2011,zhao:2012,crossfield:2012a,cowan:2012,crossfield:2012d,swain:2012}. Analysis
of the earlier observations indicated an unusual atmospheric carbon to
oxygen (C/O) ratio greater than one \citep{madhusudhan:2011}.

However, the validity of the high C/O conclusion is challenged by new
4.5\,\micron\ data \citep{cowan:2012}, the discovery of an M dwarf
only 1'' from WASP-12 which has contaminated past measurements
\citep{bergfors:2013}, and our new 2.315\,\micron\ narrowband
occultation measurement \citep{crossfield:2012d}.  Furthermore, a new
controversy has arisen: recent 1--1.7\,\micron\ spectroscopy has been
claimed to either support the high C/O model \citep{madhusudhan:2012b}
or to be inconsistent with it and to instead indicate a
hydride-dominated atmosphere \citep{swain:2012}.

\section{WASP-12b's Emission Spectrum}
\label{sec:cont}

\subsection{Photometric Measurements}
As noted above, many secondary eclipses of WASP-12b have been observed
using broadband photometry; the measurements span from
0.8--8\,\micron\ \citep{lopez-morales:2010, croll:2011a,
  campo:2011,zhao:2012,cowan:2012}. Most of these measurements are
consistent with the high C/O interpretation, with one important
exception. An analysis of WASP-12b's 4.5\,\micron\ phase curve and
eclipses \citep{cowan:2012} reported two possible conclusions: either
the planet's secondary eclipse is significantly deeper at this
wavelength than previously reported \citep{campo:2011}, or the
occultations are consistent but the planet is extremely prolate (with
an aspect ratio of 1.8).  The latter possibility has been ruled out by
recent WFC3 occultation spectroscopy (\citep{swain:2012}; see also
below), so the planet must be brighter at 4.5\,\micron\ than
previously thought.

\begin{figure*}[b!]
\centering
\includegraphics[height=5cm]{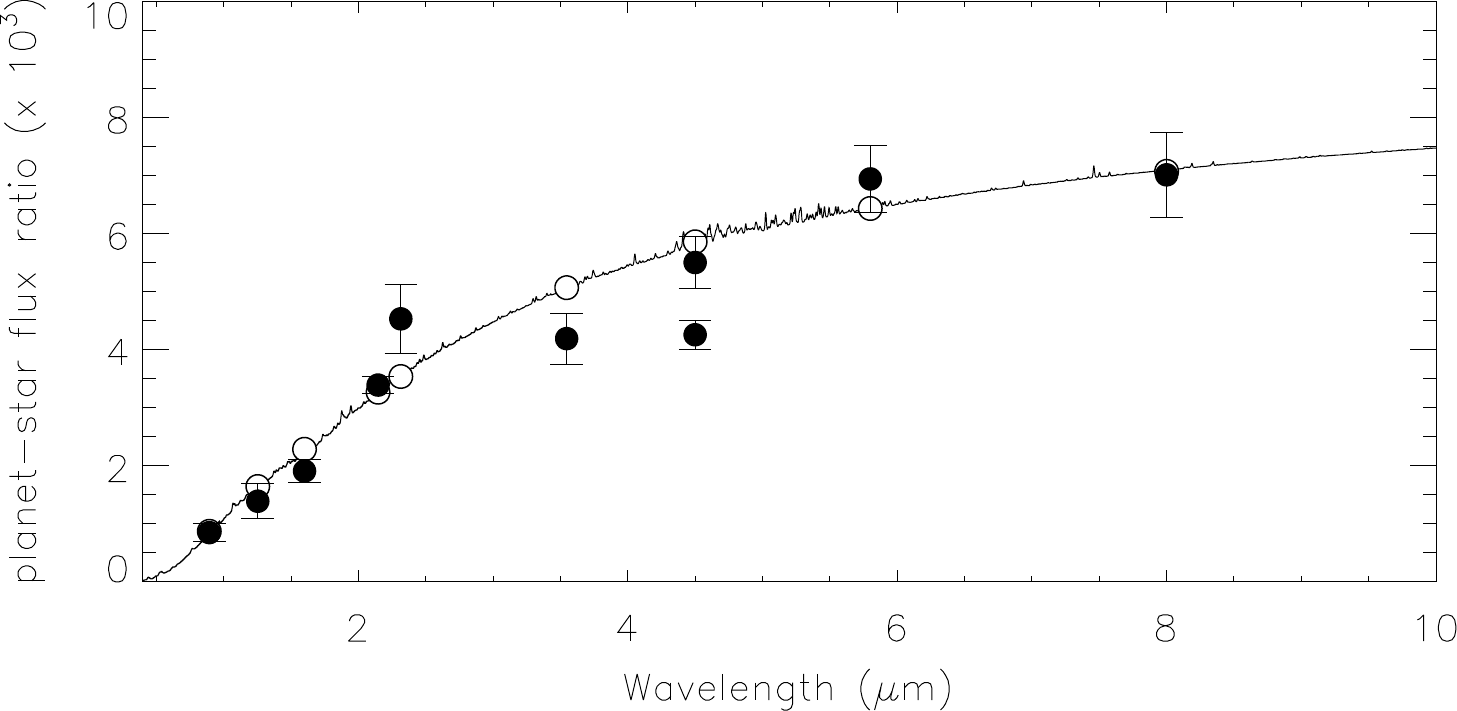}
\caption{WASP-12b broad band emission spectrum (from
  \citep{crossfield:2012d}). Solid points are dilution-corrected
  photometric secondary eclipse depths; at 4.5\,\micron\ we plot the
  two discrepant eclipse depths \citep{campo:2011,cowan:2012}. The
  solid line shows the best-fit blackbody model; open symbols show
  band-integrated model values.}
\label{fig:phot}       
\end{figure*}

In addition, our recent 2.315\,\micron\ narrowband photometry
\citep{crossfield:2012d} showed that the planet is also much brighter
at this wavelength (by $>3 \sigma$) than high C/O models predict
\citep{madhusudhan:2011}. We show the full photometric emission
spectrum of WASP-12b in Fig.~\ref{fig:phot} (after correcting for
Bergfors-6's dilution; see \citep{crossfield:2012d}). The impression
given is that of a blackbody-like spectrum with few or no strong
features.  A simple blackbody model (representing a nearly isothermal
dayside photosphere) gives $\chi^2=25$ and BIC\footnote{Bayesian
  Information Criterion (BIC) = $\chi^2 + k \ln N$, where $k$ is the
  number of free parameters and $N$ the number of data points. When
  comparing two models to a data set, the model giving the lower BIC
  is statistically preferred.}~$=29$ The initial model used to claim a
high C/O ratio gave $\chi^2=10$ and BIC~$=32$
\citep{madhusudhan:2011}; these values increase by $\sim$20 when
including the 2.315\,\micron\ and average 4.5\,\micron\ measurements.

Thus WASP-12b's photometric emission spectrum can be adequately
explained by a nearly isothermal photosphere. In this case, the data
do not justify claims of a high C/O ratio in WASP-12b's atmosphere. We
note here that a third 4.5\,\micron\ occultation has been observed,
which is more consistent with the shallow measurement than the deeper
result (J. Harrington, private communication); if confirmed, this
would have important repercussions for the interpretation of the
planet's spectrum.  A thorough and homogeneous analysis of all
available WASP-12b occultations would be of great utility.

\subsection{Near-infrared 1--1.7\,\micron\ Spectroscopy}
\vspace{-0.15cm}
Recently, high-precision spectroscopic observations of WASP-12b's NIR
emission have been obtained with HST/WFC3 \citep{swain:2012}. This
analysis notes that the initial C-rich model can match ``the slope,
but not the modulation'' of their new near-infrared spectrum and that
``the detailed shape of the emission spectrum [cannot] be modeled
using CO/CO$_2$/CH$_4$ opacity.''  Instead, the authors suggest that
an atmosphere with a significant hydride complement better explains
the NIR spectrum \citep{swain:2012}.  However, an independent
reanalysis of this emission spectrum states that the authors of the
spectroscopic analysis above ``suggested that ... models with C/O~$\ge
1$ best explained their data'' \citep{madhusudhan:2012b}, which seems
to contradict the previously quoted statement.

\begin{figure*}[b!]
\centering
\includegraphics[height=5.25cm]{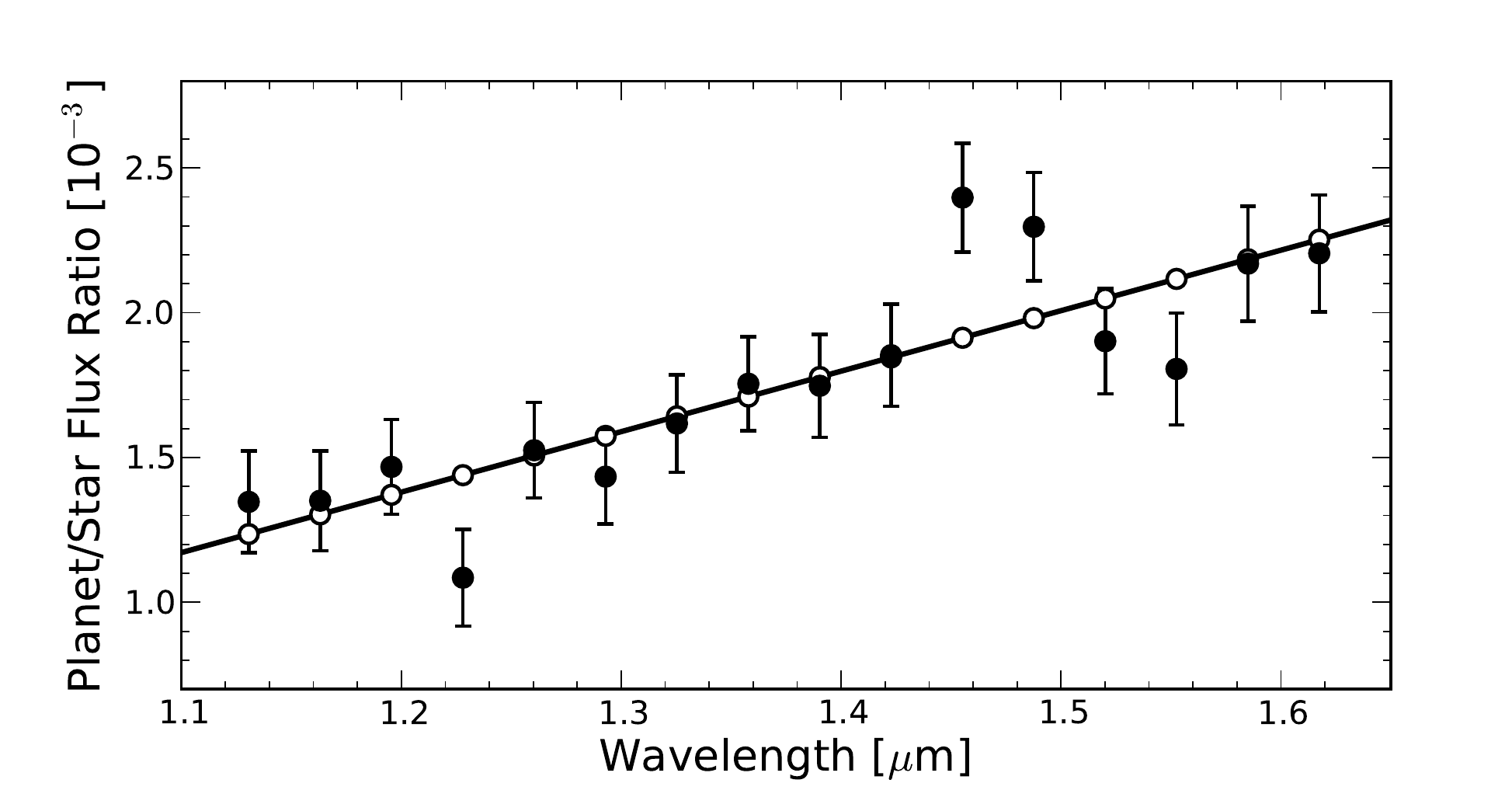}
\caption{WASP-12b spectroscopic emission spectrum. Solid points show
  the HST/WFC3 spectroscopic measurements \citep{swain:2012} after
  correction for Bergfors-6's dilution \citep{crossfield:2012d}.  The
  solid line shows the weighted best-fit linear model; open symbols
  show band-integrated model values. }
\label{fig:spec}       
\end{figure*}

Both these analyses probably read too much into the available data.
We correct the 1.0--1.7\,\micron\ emission spectrum for  eclipse
dilution  (following \citep{crossfield:2012d})
and show it in Fig.~\ref{fig:spec}. We also show a simple weighted
linear fit to the data, which gives $\chi^2=19$ and BIC~$=25$.  The
hydride-dominated model gives $\chi^2$ of roughly 13
\citep{swain:2012}; assuming $k=10$ this gives BIC~$=41$,
significantly larger than for our simpler model. Similarly, the
updated high C/O model gives $\chi^2=22$ and BIC~$=50$ for these data
\citep{madhusudhan:2012b}; even if this model were fit only to the new
WFC3 data, such a fit must give BIC~$>k \ln N=28$ and so would still
not be justified over the featureless toy model shown in
Fig.~\ref{fig:spec}.

One conclusion does seem robust: WASP-12b's 1.0--1.7\,\micron\
emission spectrum shows no evidence for emission or absorption by
water \citep{swain:2012,madhusudhan:2012b}.  This result can be
attributed either to a paucity of water in the planet's dayside
atmosphere or to a dayside photosphere which is nearly isothermal.

\vspace{-0.25cm}
\section{Conclusions}
\vspace{-0.25cm}
At present, no strong conclusions can be drawn about the composition
of WASP-12b's atmosphere on the basis of dayside (occultation)
emission measurements. If future occultation measurements robustly
detect strong emission or absorption by any species, this result would
indicate a non-isothermal dayside photosphere; the lack of water
features in WASP-12b's spectrum could then be attributed to a high C/O
ratio. The detection of only weak (or no) spectral features would be
consistent with a roughly isothermal dayside photosphere, in which
case secondary eclipse observations will be poorly suited to constrain
the planet's atmospheric makeup. In this case, observations during
primary transit may offer the best opportunity to determine the
atmospheric composition of this unusual planet.


\begin{thebibliography}{22}

\bibitem{howard:2010}
A.W. {Howard}, G.W. {Marcy}, J.A. {Johnson}, D.A. {Fischer}, J.T. {Wright},
  H.~{Isaacson}, J.A. {Valenti}, J.~{Anderson}, D.N.C. {Lin}, S.~{Ida}, Science
  \textbf{330}, 653 (2010), \texttt{1011.0143}

\bibitem{howard:2012}
A.W. {Howard}, G.W. {Marcy}, S.T. {Bryson}, J.M. {Jenkins}, J.F. {Rowe}, N.M.
  {Batalha}, W.J. {Borucki}, D.G. {Koch}, E.W. {Dunham}, T.N. {Gautier}, III
  et~al., \apjs \textbf{201}, 15 (2012), \texttt{1103.2541}

\bibitem{henry:2000}
G.W. {Henry}, G.W. {Marcy}, R.P. {Butler}, S.S. {Vogt}, \apjl \textbf{529}, L41
  (2000)

\bibitem{charbonneau:2000}
D.~{Charbonneau}, T.M. {Brown}, D.W. {Latham}, M.~{Mayor}, \apjl \textbf{529},
  L45 (2000), \texttt{arXiv:astro-ph/9911436}

\bibitem{charbonneau:2005}
D.~{Charbonneau}, L.E. {Allen}, S.T. {Megeath}, G.~{Torres}, R.~{Alonso}, T.M.
  {Brown}, R.L. {Gilliland}, D.W. {Latham}, G.~{Mandushev}, F.T. {O'Donovan}
  et~al., \apj \textbf{626}, 523 (2005), \texttt{arXiv:astro-ph/0503457}

\bibitem{deming:2005}
D.~{Deming}, T.M. {Brown}, D.~{Charbonneau}, J.~{Harrington}, L.J.
  {Richardson}, \apj \textbf{622}, 1149 (2005), \texttt{arXiv:astro-ph/0412436}

\bibitem{harrington:2006}
J.~{Harrington}, B.M. {Hansen}, S.H. {Luszcz}, S.~{Seager}, D.~{Deming},
  K.~{Menou}, J.~{Cho}, L.J. {Richardson}, Science \textbf{314}, 623 (2006)

\bibitem{knutson:2007b}
H.A. {Knutson}, D.~{Charbonneau}, L.E. {Allen}, J.J. {Fortney}, E.~{Agol}, N.B.
  {Cowan}, A.P. {Showman}, C.S. {Cooper}, S.T. {Megeath}, \nat \textbf{447},
  183 (2007), \texttt{0705.0993}

\bibitem{hebb:2009}
L.~{Hebb}, A.~{Collier-Cameron}, B.~{Loeillet}, D.~{Pollacco},
  G.~{H{\'e}brard}, R.A. {Street}, F.~{Bouchy}, H.C. {Stempels}, C.~{Moutou},
  E.~{Simpson} et~al., \apj \textbf{693}, 1920 (2009), \texttt{0812.3240}

\bibitem{chan:2011}
T.~{Chan}, M.~{Ingemyr}, J.N. {Winn}, M.J. {Holman}, R.~{Sanchis-Ojeda},
  G.~{Esquerdo}, M.~{Everett}, \aj \textbf{141}, 179 (2011), \texttt{1103.3078}

\bibitem{maciejewski:2011}
G.~{Maciejewski}, R.~{Errmann}, S.~{Raetz}, M.~{Seeliger}, I.~{Spaleniak},
  R.~{Neuh{\"a}user}, \aap \textbf{528}, A65 (2011), \texttt{1102.2421}

\bibitem{croll:2011a}
B.~{Croll}, D.~{Lafreniere}, L.~{Albert}, R.~{Jayawardhana}, J.J. {Fortney},
  N.~{Murray}, \aj \textbf{141}, 30 (2011), \texttt{1009.0071}

\bibitem{campo:2011}
C.J. {Campo}, J.~{Harrington}, R.A. {Hardy}, K.B. {Stevenson}, S.~{Nymeyer},
  D.~{Ragozzine}, N.B. {Lust}, D.R. {Anderson}, A.~{Collier-Cameron},
  J.~{Blecic} et~al., \apj \textbf{727}, 125 (2011), \texttt{1003.2763}

\bibitem{cowan:2012}
N.B. {Cowan}, P.~{Machalek}, B.~{Croll}, L.M. {Shekhtman}, A.~{Burrows},
  D.~{Deming}, T.~{Greene}, J.L. {Hora}, \apj \textbf{747}, 82 (2012),
  \texttt{1112.0574}

\bibitem{crossfield:2012d}
I.J.M. {Crossfield}, T.~{Barman}, B.M.S. {Hansen}, I.~{Tanaka}, T.~{Kodama},
  \apj \textbf{760}, 140 (2012), \texttt{1210.4836}

\bibitem{lopez-morales:2010}
M.~{L{\'o}pez-Morales}, J.L. {Coughlin}, D.K. {Sing}, A.~{Burrows}, D.~{Apai},
  J.C. {Rogers}, D.S. {Spiegel}, E.R. {Adams}, \apjl \textbf{716}, L36 (2010),
  \texttt{0912.2359}

\bibitem{zhao:2012}
M.~{Zhao}, J.D. {Monnier}, M.R. {Swain}, T.~{Barman}, S.~{Hinkley}, \apj
  \textbf{744}, 122 (2012), \texttt{1109.5179}

\bibitem{crossfield:2012a}
I.J.M. {Crossfield}, B.M.S. {Hansen}, T.~{Barman}, \apj \textbf{746}, 46
  (2012), \texttt{1201.1023}

\bibitem{swain:2012}
M.~Swain, P.~Deroo, G.~Tinetti, M.~Hollis, M.~Tessenyi, M.~Line, H.~Kawahara,
  Y.~Fujii, A.~Showman, S.~Yurchenko, arXiv.org \textbf{astro-ph.EP} (2012),
  \texttt{1205.4736}

\bibitem{madhusudhan:2011}
N.~{Madhusudhan}, J.~{Harrington}, K.B. {Stevenson}, S.~{Nymeyer}, C.J.
  {Campo}, P.J. {Wheatley}, D.~{Deming}, J.~{Blecic}, R.A. {Hardy}, N.B. {Lust}
  et~al., \nat \textbf{469}, 64 (2011), \texttt{1012.1603}

\bibitem{bergfors:2013}
C.~{Bergfors}, W.~{Brandner}, S.~{Daemgen}, B.~{Biller}, S.~{Hippler},
  M.~{Janson}, N.~{Kudryavtseva}, K.~{Gei{\ss}ler}, T.~{Henning},
  R.~{K{\"o}hler}, \mnras \textbf{428}, 182 (2013), \texttt{1209.4087}

\bibitem{madhusudhan:2012b}
N.~{Madhusudhan}, \apj \textbf{758}, 36 (2012), \texttt{1209.2412}

\end{thebibliography}
\end{document}